\documentstyle[prb,aps]{revtex}
\begin{document}
\input{psfig}
\psfull
\tighten
\title{Inverse design of proteins with hydrophobic and polar amino acids.}
\author{Cristian Micheletti$^1$, Flavio Seno$^1$, Amos Maritan$^2$
and Jayanth R. Banavar$^3$}
\vskip 0.3cm
\address{(1)INFM-Dipartimento di Fisica, Universit\`a di Padova, Via
Marzolo 8, 35131  Padova, Italy }
\address{(2) International School for Advanced Studies (S.I.S.S.A.),
Via Beirut 2-4, 34014 Trieste, Italy }
\address{(3) Department of Physics and Center for Materials Physics,
104 Davey Laboratory, The Pennsylvania State University, University
Park, Pennsylvania 16802}
\date{\today}
\maketitle
\begin{abstract}
A two amino acid (hydrophobic and polar) scheme is used to perform the
design on target conformations corresponding to the native states of
twenty single chain proteins.  Strikingly, the percentage of
successful identification of the nature of the residues benchmarked
against naturally occurring proteins and their homologues is around 75
\% independent of the complexity of the design procedure.  Typically,
the lowest success rate occurs for residues such as alanine that have
a high secondary structure functionality.  Using a simple lattice
model, we argue that one possible shortcoming of the model studied may
involve the coarse-graining of the twenty kinds of amino acids into
just two effective types.
\end{abstract}


\noindent{{\bf PACS-classification:}  87.10.+e, 87.15By}

\section{Introduction}

The gigantic efforts spent by the scientific community over the past
decades have unravelled many of the mysteries lying behind the
chemistry and biological functionality of proteins \onlinecite{b0,b1,b2}.
However, two fundamental questions remain unanswered:

\begin{enumerate}
\item{what are the mechanisms that guide the folding of a string of
amino acids into a complicated three-dimensional structure of helices,
loops and $\beta$-sheets,}
\item{how does one identify the sequence of amino acids that folds
into a pre-assigned native structure.}
\end{enumerate}

This study, building on the pioneering work of Shakhnovich
and Gutin \onlinecite{new} and Sun {\em et al.} \onlinecite{i16}, deals with the
second issue which is commonly referred to as 
the {\em inverse folding problem} \onlinecite{i14,i10,i2,i4,i12,i15,i13,i3,i11,i8,i9}.
This problem is central in many areas of biology and medicine; indeed
the possibility of designing artificial proteins would open far-reaching
pharmaceutical applications.

\noindent The complexity of the problem is enormous because, in
principle, it entails an exhaustive comparison of the native states of
all sequences in search for the one(s) matching the desired target
structure \onlinecite{i10,i12,i11}.

This procedure has been recently formulated into a general
mathematical form appropriate for numerical implementation \onlinecite{i11}
which shows that solving the design problem for a
structure, $\Gamma$, amounts to the identification of the amino acid
sequence, $S$, that maximizes the occupation probability,
$P_\Gamma(S)$,

\begin{equation}
P_\Gamma(S) = { e^{- \beta E_\Gamma(S)} \over \sum_{\Gamma^\prime} e^{- \beta
E_{\Gamma^\prime}(S)}} = e^{ -\beta ( E_\Gamma(S) -  F(S))}
= e^{- \beta K(\Gamma, S)} \ ,
\label{eqn:occprob}
\end{equation}

\noindent where $\beta$ is the Boltzmann weight, $E_\Gamma(S)$ is the
energy of the sequence $S$ over the structure $\Gamma$ and the sum in
the denominator is taken over all possible structures, $\{
\Gamma^\prime \}$ having the same length of $\Gamma$. The winning
sequence will maximize $P_\Gamma(S)$ at all temperatures below the
folding transition temperature (where the occupation probability of
the native state is macroscopic).

The previous studies of \onlinecite{new} and Sun {\em et al.} \onlinecite{i16}
consisted of finding 
a sequence that has the lowest possible energy in the putative native state.
This is equivalent to the assumption that $F(S)$ in the above equation
is independent of $S$.
The latter study used a modified Hamiltonian that included a chemical
potential term that controlled the number of hydrophobic residues
and may be interpreted as a first approximation for the inclusion
of an $F(S)$ term (see later).

Two main obstacles need to be overcome to implement
(\ref{eqn:occprob}). The first is that one needs to know how to
calculate $E_\Gamma(S)$ and $F(S)$; second, it is necessary to explore
the whole space of sequences to find the one maximizing
(\ref{eqn:occprob}).

A commonly used simplification of the latter problem is effected on
coarse graining the 20 types of naturally occurring amino acids into
just two dominant classes: hydrophobic (H) and polar, (P). The
evidence in favour of this subdivision is considerable \onlinecite{i1,i5}

In fact, it has been shown that, to a large extent, the folding of
proteins is driven by the collapse of H residues into a compact
hydrophobic core surrounded by polar amino acids or solvent molecules
\onlinecite{i1,i5,i15a}. On top of that it appears possible to take a
protein and 
exchange some of its amino acids within the H (or P) class without
changing its native structure notably \onlinecite{i5}.

In this paper we will adopt this point of view and perform a design on
``real structures'' (i.e., structures extracted from the Protein Data
Bank) within the HP framework.  We will address the difficulty of
maximizing (\ref{eqn:occprob}) by considering a series of approximate
forms for $E_\Gamma(S)$ and $F(S)$ of increasing complexity and
refinement. For each design attempt, we present a detailed summary of
the failure rates of identifying the correct class of each amino acid.
It will appear that the lowest success is reached for residues with a
high secondary structure functionality (especially alanine). We find
that, though the failure rates on individual amino acids varies
significantly over the various design attempts, the overall design
success remains nearly constant.  This is possibly suggestive of a
limitation of the HP coarse-graining; in support of this we will
present evidence showing that the the HP coarse-graining hinders the
success of design strategies in a solvable protein model.

\section{The design algorithms}

To test the design algorithm we first chose a set of 20 single chain
proteins from the protein data bank (PDB); here and below the proteins
will be identified with their index number as in Table
\ref{tab:files}.

In order to perform the HP design, we begin by substituting each amino
acid unit with a fictitious residue placed at a distance of 3 \AA\
from the protein backbone along the $C_\alpha-C_\beta$ direction of
the true residue, following Sun {\em et al.} \onlinecite{i16}.  For GLY, which
does not have a $\beta$-Carbon, the fictitious residue coincided with
the $\alpha$-Carbon \onlinecite{i17}. Following this
procedure, one obtains the bare backbone of the original protein
stripped from any information that could distinguish different types
of amino acids.

\noindent The goal of the design procedure was to identify the
polarity or hydrophobicity of fictitious residues exactly as in the
true sequence, where the 20 aminoacids were coarse-grained as in Table
\ref{tab:hp}.

{}From table \ref{tab:hp} it can be calculated that the fraction of H
residues is about $41\%$. If one were to use a totally random method
for guessing the correct class of residues, while respecting the H/P
ratio $r = 0.41/0.59$ one would obtain a relative success of

\begin{equation}
{1 \over 2} + {1 \over 2} \biggl( {r -1 \over r+1} \biggr)^2 \approx 52\%
\end{equation}
\noindent only slightly higher than without constraining $r$. This
result is to be borne in mind when assessing the performance of the
design procedures presented in the remainder of this section.

\subsection{Method 1}

The coarsest design attempt that can be tried is to establish the
polarity of a residue according to the number of neighbours in contact
with it. Customarily two non-consecutive residues, $i$ and $j$, are
said to be in contact if their distance, $r_{ij}$, falls within a
range of around $7 \AA$\ and no other residue is between them. To each
residue, $i$, in the target structure we assigned a contact score
according to the rule

\begin{equation}
n_c(i) = \tilde{\sum_{j}} {1 \over 1 + e^{r_{ij} -6.5}},
\label{eqn:weight}
\end{equation}

\noindent which weights the strength of contact interactions with a
smooth sigmoidal function \onlinecite{i16,i17}. The superscript tilde in
equation (\ref{eqn:weight}) 
indicates that the sum is not taken over $j \in \{i-1, i, i+1\}$, nor
over residues that are spaced more than $10 \AA$ apart. The latter
constraint is used to avoid the possibility that an intervening
residue is between sites $i$ and $j$. The residues which
have a contact score greater than 5 were chosen to belong to the H
class while the others were considered polar (P). Finally, the
obtained string was compared with the true (HP-coarse grained) protein
sequence. It turned out that, on average, the design procedure
identified the correct class of amino acid residues 73\% of the
time. This very simple design strategy matches the performance of
previous design attempts which were based on more complex algorithms
where, for example, the polarity of residues was assigned according to
the exposed area of their Van der Waals spheres (Sun {\em et al.} \onlinecite{i16}.

\begin{figure}[htbp]
\centerline{\psfig{figure=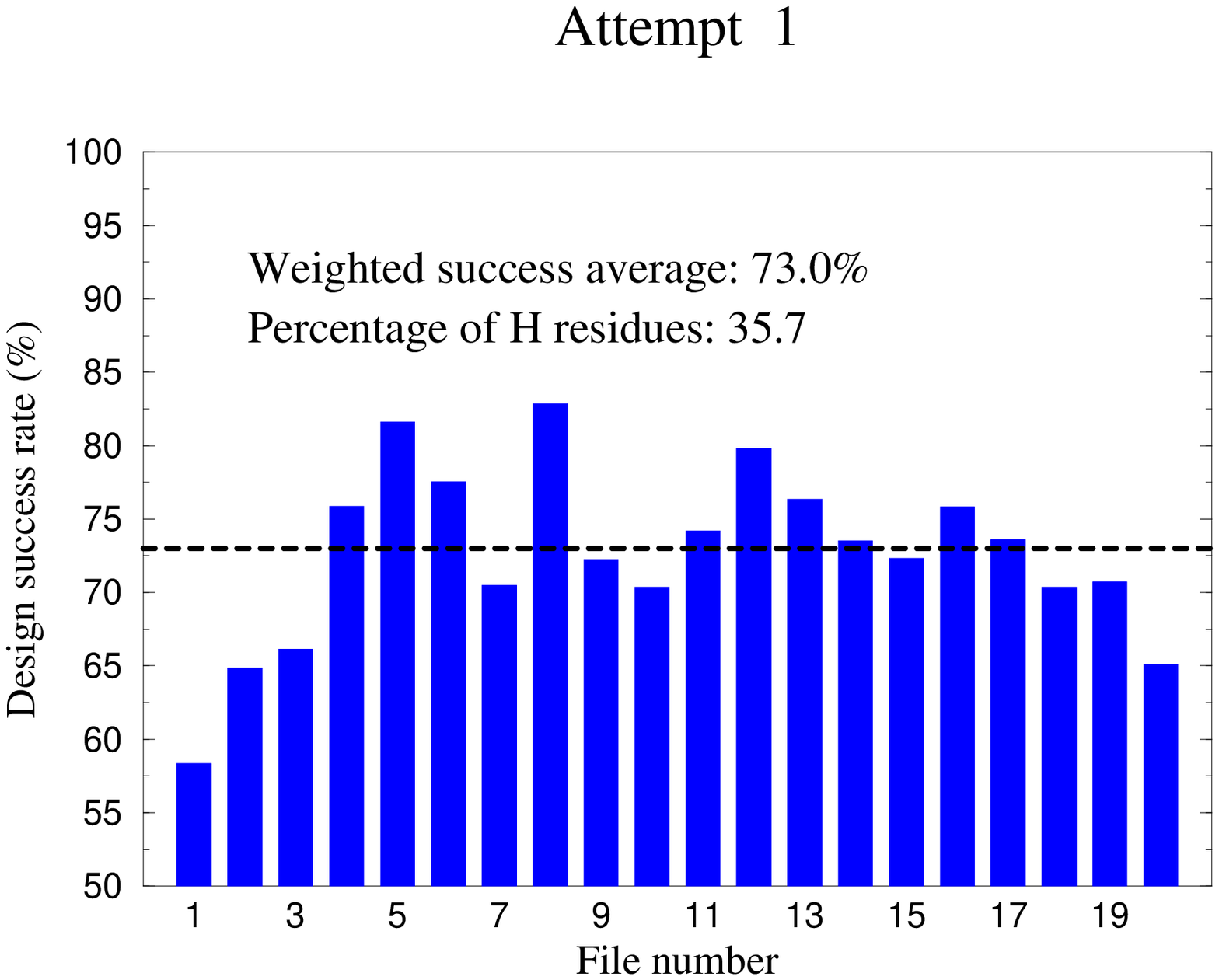,width=3.5in}}
\vskip 0.5cm
\caption{Histogram of the design success rates on each of the 20
proteins of Table \protect{\ref{tab:files}} obtained using method
1.}
\label{fig:a}
\end{figure}

\begin{figure}[htbp]
\centerline{\psfig{figure=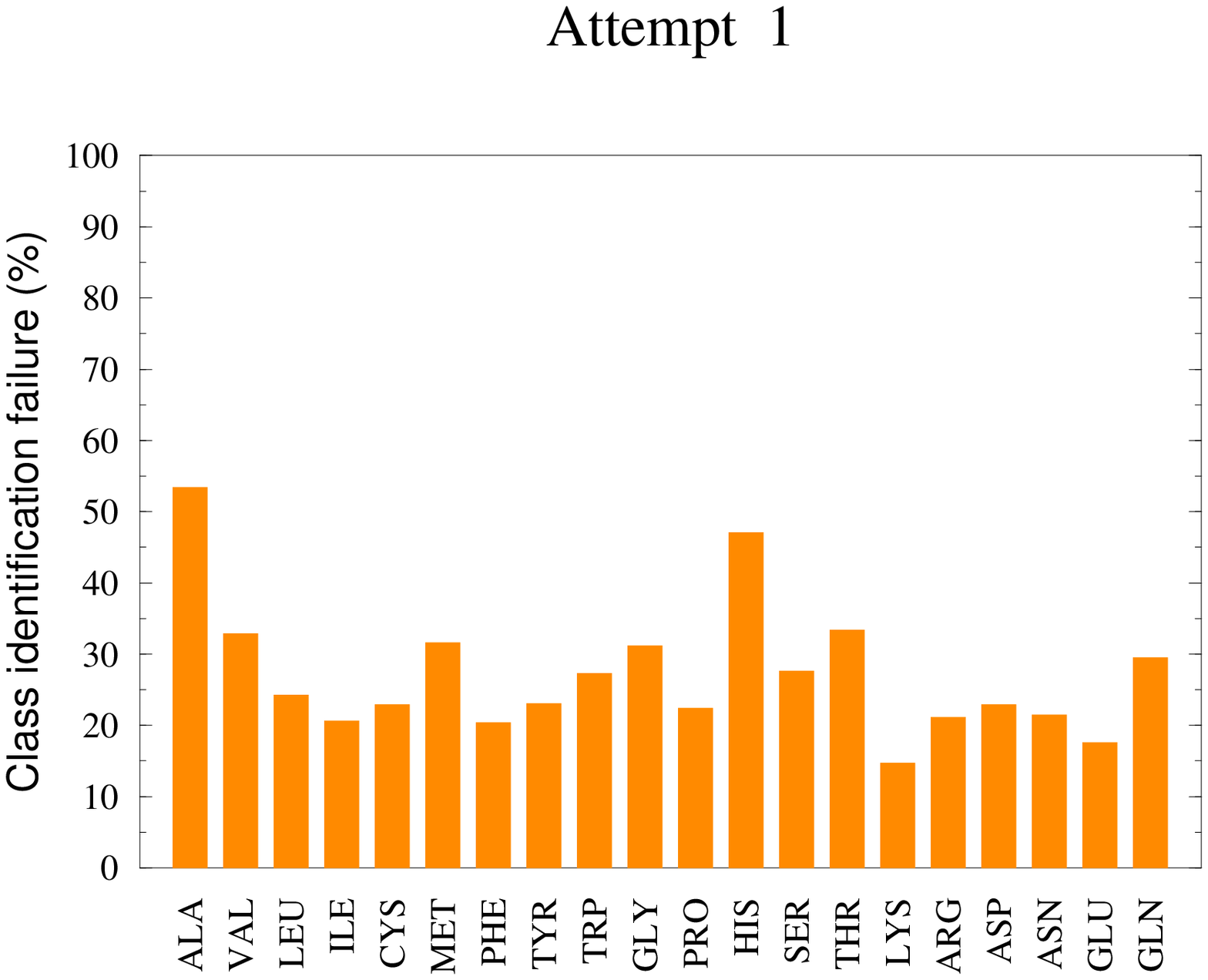,width=3.5in}}
\vskip 0.5cm
\caption{Histogram of the failure rates in identifying the correct H/P
class of the 20 amino acids obtained with method 1.}
\label{fig:b}
\end{figure}

The detailed success rate on each protein in our chosen set is given
in Fig.~\ref{fig:a}. Fig.~\ref{fig:b}, on the other hand, shows the
frequency with which a given amino acid was assigned to the wrong HP
class. These results, which to our knowledge were not considered in
previous studies, provide useful insight for the design failure. For
example, from Fig.~\ref{fig:b}, it appears that the class
identification failure is highest for alanine, for which it is
slightly higher than 50\%. This can probably be ascribed to the fact
that the sites occupied by alanine are determined more on the basis of
steric interactions than hydrophobic ones. The failure of identifying
alanine as hydrophobic seriously affects the overall design success
rate; in fact, nearly 10\% of proteins residues are of this type (see
Table \ref{tab:hp}).

\subsection{Method 2}

It is perhaps surprising that the previous design method can yield a
success rate of 73\% especially in view of the fact that it only uses
local geometrical information about the target structure, $\Gamma$. In
this subsection, we go beyond this approximation and take as the
solution 
to the design problem, the sequence, $S$ minimizing the following
form (see also Sun {\em et al.} \onlinecite{i16}) 
for $K(\Gamma,S)$ appearing in (\ref{eqn:occprob}),

\begin{equation} K_\Gamma(\Gamma,S) = \sum_{i,j} \epsilon(S_i, S_j)
f(r_{ij}) + \mu \sum_i S_i
\label{eqn:en2}
\end{equation}

\noindent where $S_i =0$ [$S_i =1$] if residue $i$ is of type P [H],
$\epsilon$ is the contact energy matrix, $f(r_{ij})$ is the sigmoidal
contact-strength function,

\begin{equation}
f(r_{ij}) = \left\{
\begin{array}{l l}
{(1 + e^{r_{ij} -6.5})^{-1}} & \ \mbox{if $j \not\in
\{i-1,i,i+1\}$\ and $r_{ij} < 10 \AA$,}\\
0 & \mbox{otherwise,}
\end{array}
\right.
\end{equation}

\noindent and $\mu$ is a positive quantity. The contact matrix,
$\epsilon$ is chosen to be symmetric (i.e., $\epsilon(0,1)=
\epsilon(1,0)$); moreover the energy scale is set so that
$\epsilon(1,1)=-1$. The second term in (\ref{eqn:en2}) does not depend
on the target structure, $\Gamma$, and is to be regarded as an
approximate expression for the free-energy , $F(S)$, whose effect is
to control the ratio of $H$ to $P$ residues in designed
sequences. This particular approximation for $F$ was inspired by the
fact that, on HP lattice models, sequences with the same number of $H$
and $P$ residues have nearly the same free-energy \onlinecite{i18}. The
introduction of the "chemical potential", $\mu$, 
appears the simplest way of sifting through the sequence space to
retain only sequences with the designed H-P ratio. Among this subset,
the putative solution to the design problem will be the sequence
minimizing the contact energy term. On the basis of the success of an
analogous HP design strategy for three-dimensional lattice structures
\onlinecite{i18}, it can be expected that the sequence
minimizing (\ref{eqn:en2}) will be close to the true protein sequence.

The quantities $\epsilon(1,0)$, $\epsilon(0,0)$ and $\mu$ appearing in
(\ref{eqn:en2}) are regarded as parameters to be optimized in order to
maximize the overall design success rate. This step can be viewed as a
way of extracting the HP contact energies from proteins of known
sequence and conformation.

The design procedure engine consists of the following two steps:

\begin{enumerate}
\item{for a given set of parameters the sequence $S$ minimizing
(\ref{eqn:en2}) is identified with a simulated annealing procedure
(the elementary move being the mutation of a fraction of residues
from one class to the other),}

\item{the parameters are varied, and step 1 is repeated, in the
attempt to identify the set of values giving the highest average
design success rate.}
\end{enumerate}

\begin{figure}[htbp]
\centerline{\psfig{figure=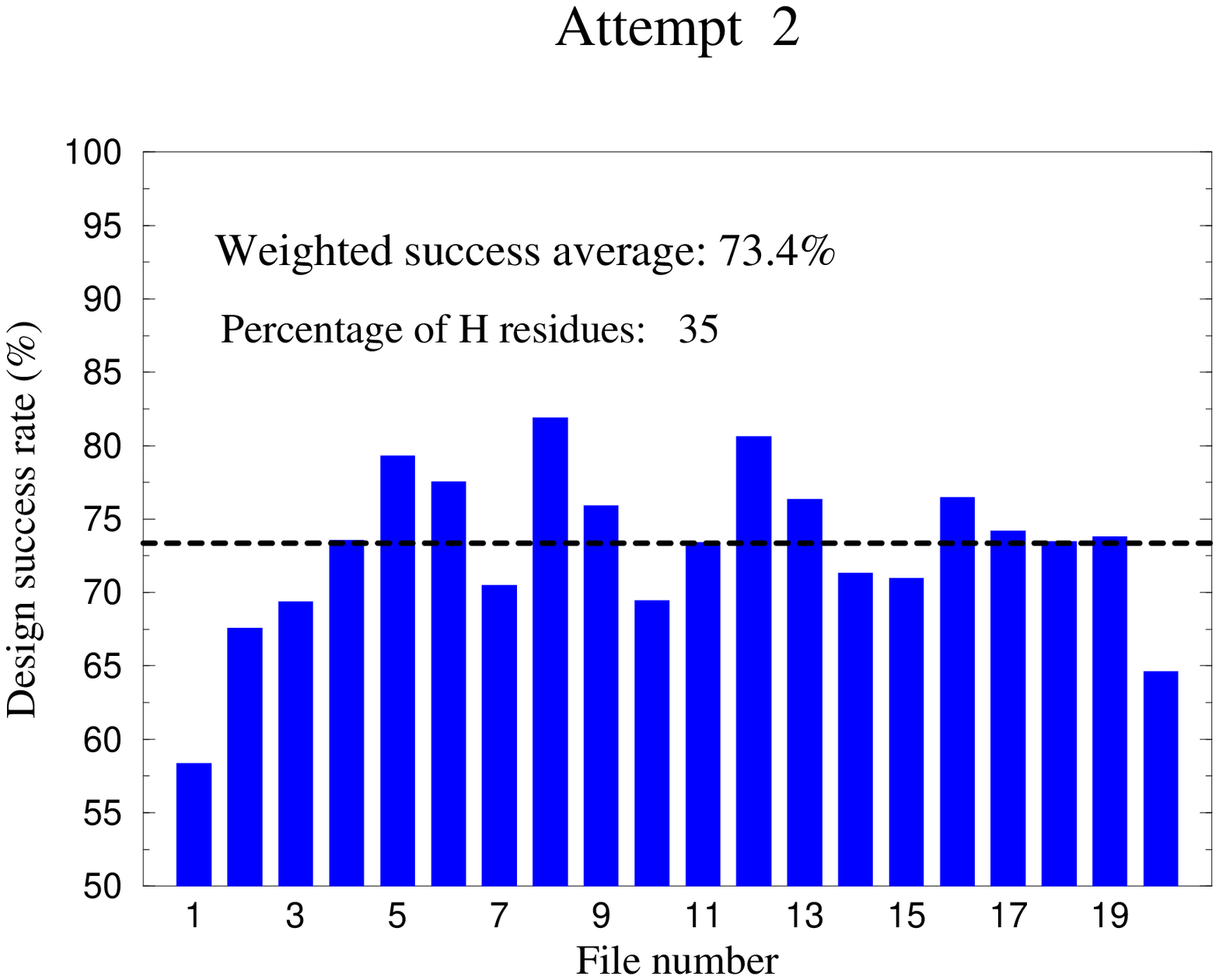,width=3.5in}}
\vskip 0.5cm
\caption{Histogram of the design success rates on each of the 20
proteins of Table \protect{\ref{tab:files}} obtained using method
2.}
\label{fig:c}
\end{figure}

\noindent The highest success rate was found  for
\begin{equation}
\epsilon(0,1)= 0.006025 \ , \epsilon(0,0)=1.481606 \ , \mu =
6.1909
\end{equation}
\noindent and was equal to 73.4\% as shown in Fig.~\ref{fig:c}.

\begin{figure}[htbp]
\centerline{\psfig{figure=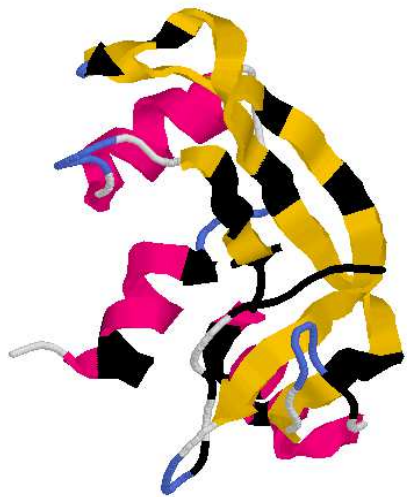,width=3.5in}}
\vskip 0.5cm
\caption{Ribbon plot of protein 3rn3. The black sections highlight
residues which were wrongly assigned by the second design procedure.}
\label{fig:3rn3}
\end{figure}

Figure \ref{fig:3rn3} represents the ribbon plot of protein 3rn3,
where the design success was 73.3\%. Helices and $\beta-sheets$ are
coloured in purple and yellow respectively, while the black portions
mark the residues whose hydrophobicity/polarity was not recognized
correctly by the design method.

\begin{figure}[htbp]
\centerline{\psfig{figure=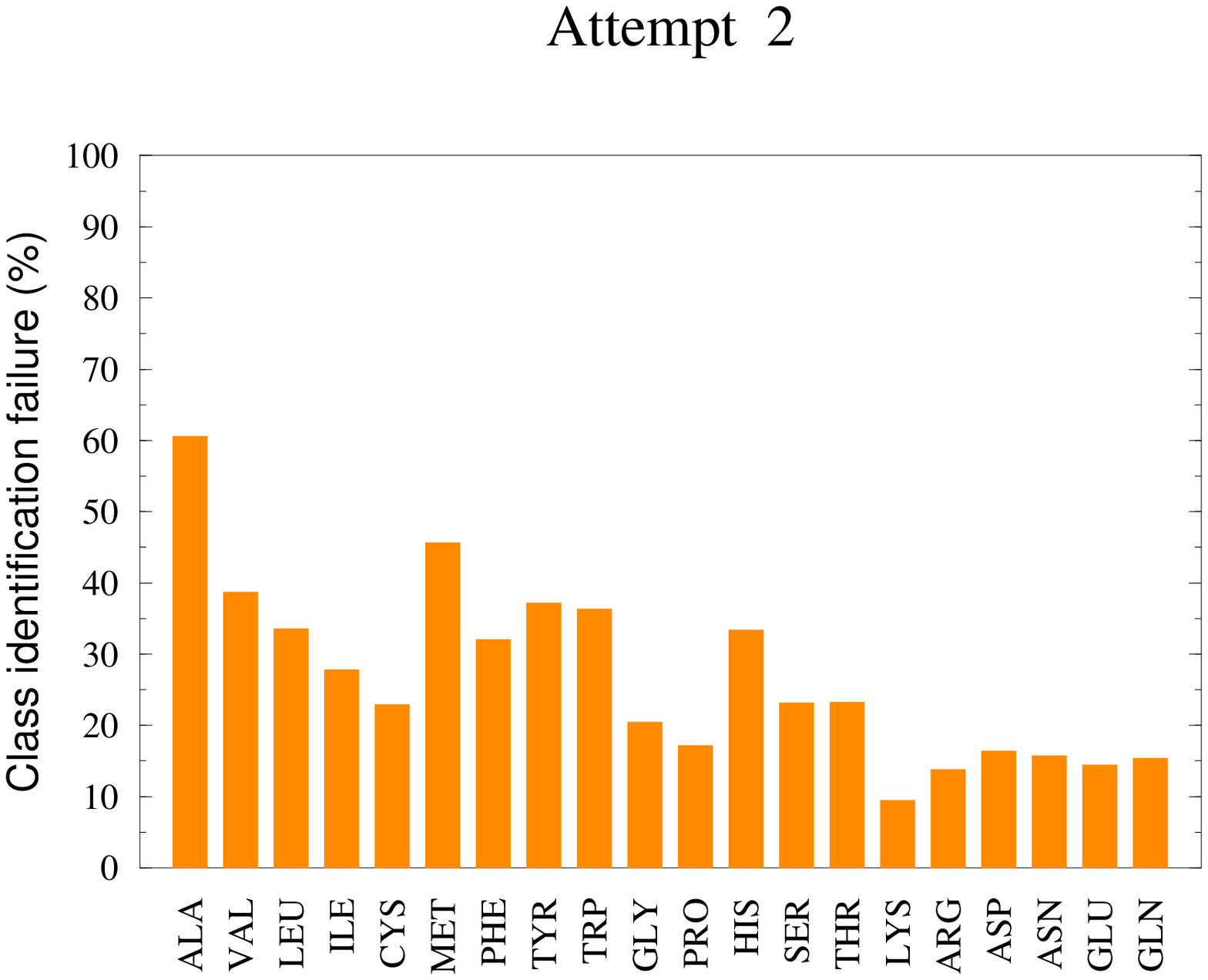,width=3.5in}}
\vskip 0.5cm
\caption{Histogram of the failure rates in identifying the correct H/P
class of the 20 amino acids obtained with method 2.}
\label{fig:d}
\end{figure}

Remarkably, despite the increased computational effort, the overall
success rate has not improved appreciably over the previous design
attempt. In particular it can be noted that, while the failure rate
over the individual files is roughly the same as in Fig.~\ref{fig:a},
Figs.  \ref{fig:b} and \ref{fig:d} differ significantly. The average
failure in identifying P-type residues has decreased significantly
while the failure rate for alanine has grown from 50\% to 60\%. As
mentioned before this can be ascribed to the small volume of alanine
which favours its location in protein structures on the basis of
excluded volume reasons. A high failure rate also affects metionine,
which is a common helix-former. Hence, it appears that the highest
failure rate is found among residues whose location in the protein is
dictated by specific functionality rather than mere energetic
considerations.

\subsection{Method 3}

In a third attempt we tried to improve the approximation for the free
energy term, $F(S)$, used in (\ref{eqn:en2}). A possible way of
constructing approximate forms for $F(S)$ is to include in it all
possible constraints that are satisfied by real protein sequences. One
of these constraints is that, using the coarse-graining scheme of
Table \ref{tab:hp}, the ratio of $H$ to $P$ residues must be,

\begin{equation}
r \approx 0.71 \ .
\label{eqn:constr1}
\end{equation}

\noindent The previous method tried to tune in to the right value of
$r$ by optimizing the ``chemical potential'', $\mu$.

\begin{figure}[htbp]
\centerline{\psfig{figure=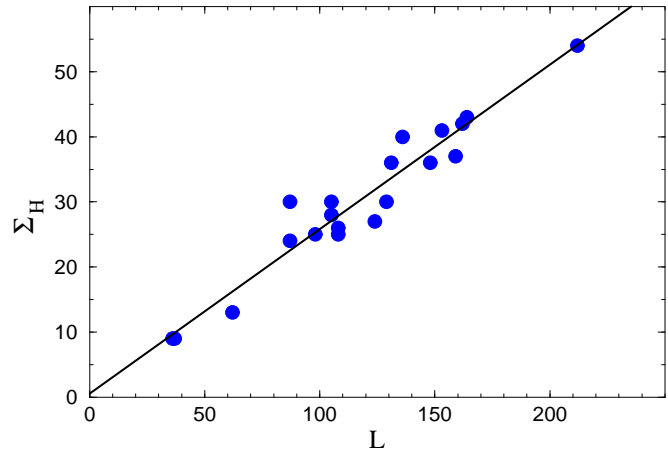,width=3.5in}}
\vskip 0.5cm
\caption{Plot of the number of H segments, $\Sigma_H$ for our set of 20
proteins as a function of the protein length, $L$. The solid curve
is the interpolating line (see equation
\protect{\ref{eqn:constr2}}).}
\label{fig:sigma}
\end{figure}

By studying the statistical properties of protein chains we have been
able to identify another constraint which, to our knowledge, had not
been previously identified. In fact, we have established that the
number of $H$ segments, $\Sigma_H$, in a variety of proteins grows
linearly with the length, $L$, of the chain. As can be seen in
Fig.~\ref{fig:sigma} the linear behaviour is marked and, in fact, the
linear correlation coefficient over our set of 20 proteins was equal
to $0.964$.

\noindent A linear regression of the points in Fig.~\ref{fig:sigma}
gives the following equation for the interpolating line:

\begin{equation}
\tilde{\Sigma}_H(L) =0.547889 + 0.252676  \cdot L
\label{eqn:constr2}
\end{equation}

\noindent For chains of length $L \approx 130$ the difference between
the true value of $\Sigma_H$ ($\approx 35$) and the one estimated with
(\ref{eqn:constr2}) was of the order of 3 units.  Given the high
degree of reliablity of constraint (\ref{eqn:constr2}) we decided to
incorporate it in our expression for $F(S)$.  In fact, it turned out
that the sequences designed with the previous method had rather low
values of $\Sigma_H$; in other words the $H$ residues tended to
cluster together in relatively long segments.

Our expectation was that the simultaneous requirement that design
sequences should obey both (\ref{eqn:constr1}) and (\ref{eqn:constr2})
would be an efficient sieve for isolating good sequences. Hence we
adopted the following form for $K(\Gamma, S)$,

\begin{eqnarray}
&& K_\Gamma(\Gamma,S) = \sum_{i \not= j} \epsilon(S_i, S_j) f(r_{ij})
+\nonumber \\
&& \biggl[ V_1( \Sigma_H(S) - \tilde{\Sigma}_H(L) )^2 +\nonumber \\
&&V_2 (n_H(S) - 0.415 \cdot L)^2 \biggr]
\label{eqn:en3}
\end{eqnarray}

\noindent where $L$, $N_H(S)$ and $\Sigma_H(S)$ are, respectively, the
length of the chain, the number of $H$ residues in $S$ and the number
of $H$ segments in $S$. The term in square bracket in equation
(\ref{eqn:en3}) can be regarded as an expansion of the free energy,
$F(S)$ (see eqn. (\ref{eqn:occprob})), to fourth order in the spin
variables, $S_i$. In fact, it can be equivalently recast into,

\begin{equation}
{V_1 \over 2} \sum_{i=1}^{L-1} \biggl[ (S_i - S_{i+1})^2 + S_0 + S_L - 
\tilde{\Sigma}_H(L) \biggr]^2 + V_2 \sum_{i=1}^L \biggl( S_i - {r \over
r+1} \biggr)^2.
\end{equation}

\begin{figure}[htbp]
\centerline{\psfig{figure=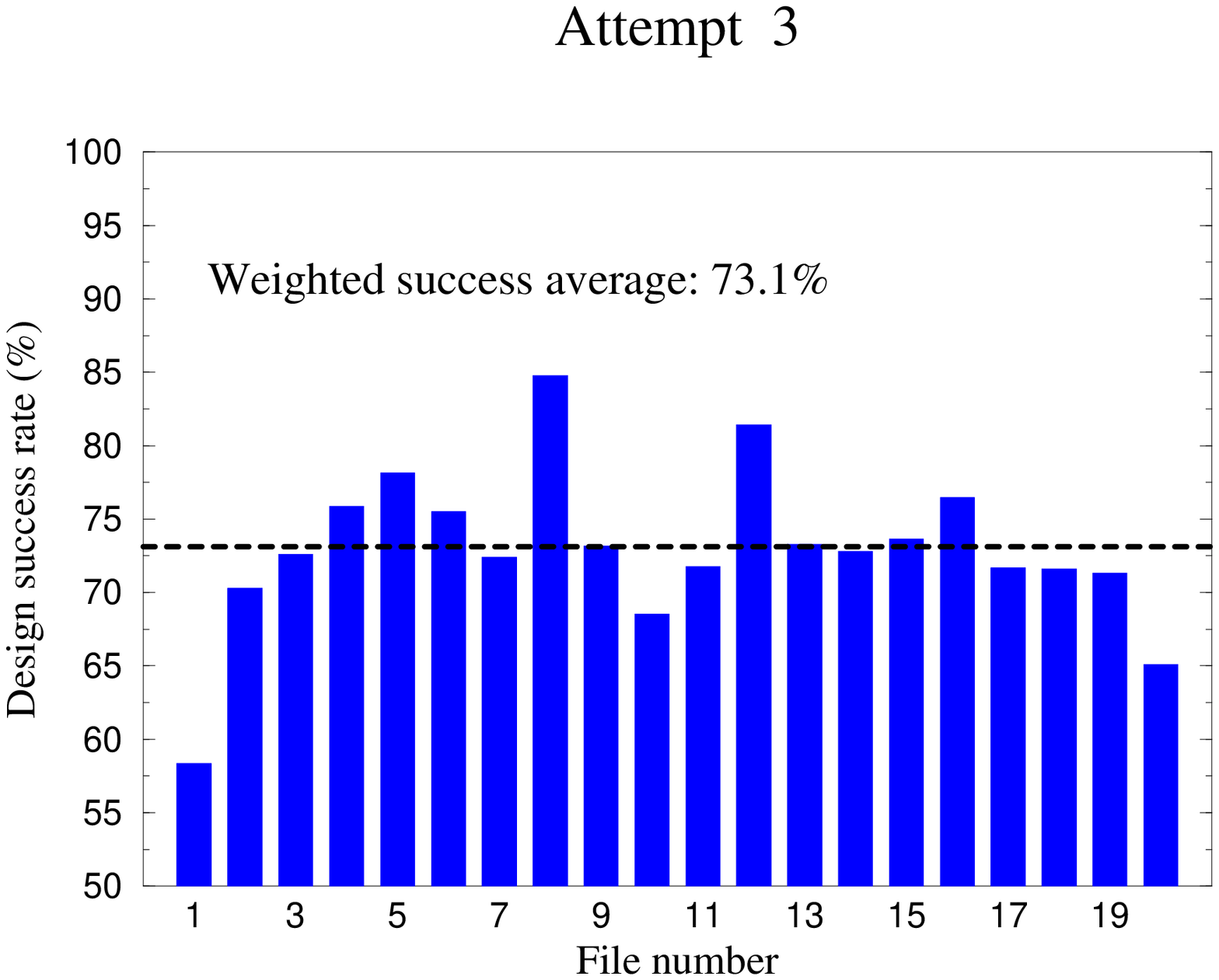,width=3.5in}}
\vskip 0.5cm
\caption{Histogram of the design success rates on each of the 20
proteins of Table \protect{\ref{tab:files}} obtained using method
3.}
\label{fig:e}
\end{figure}

\begin{figure}[htbp]
\centerline{\psfig{figure=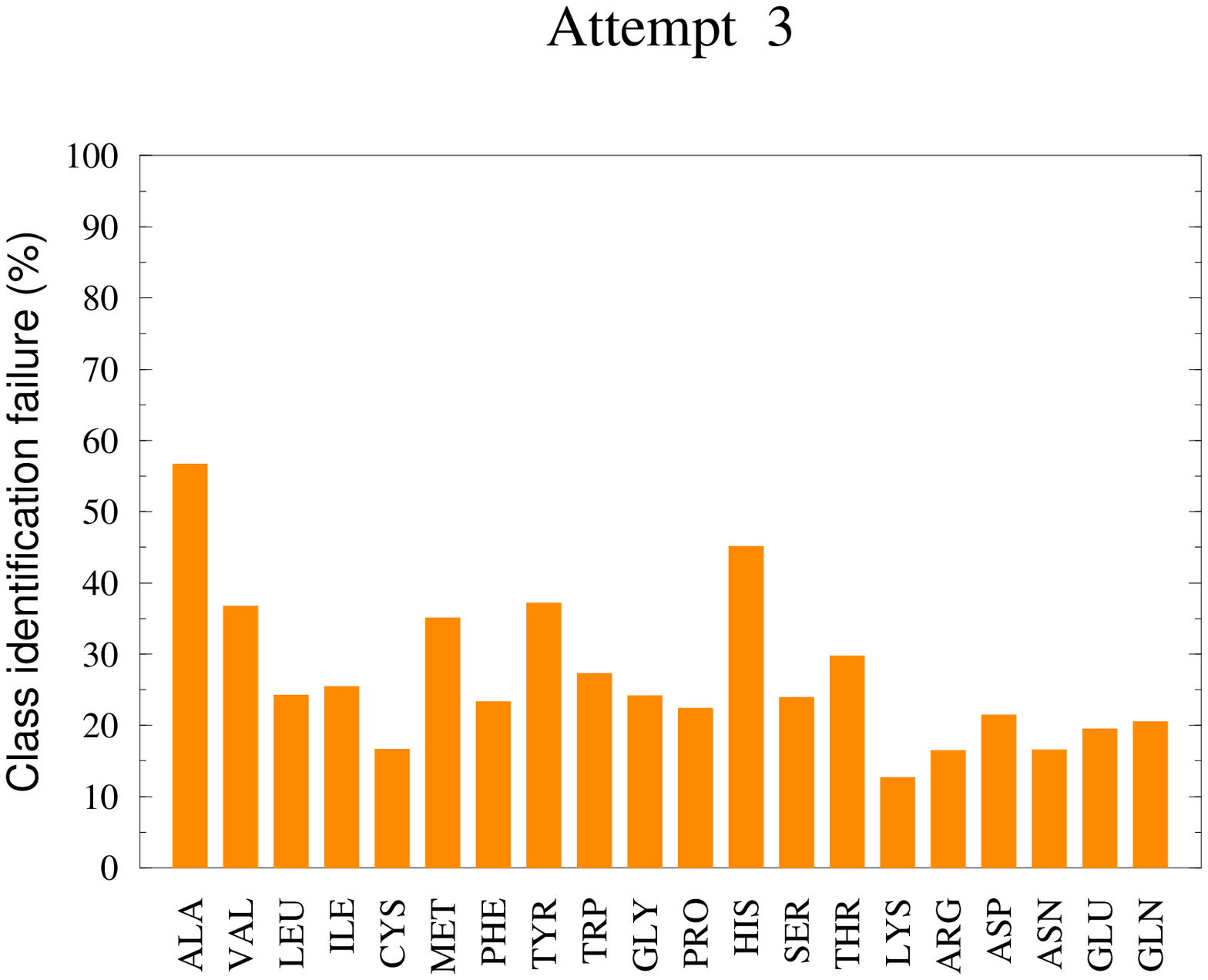,width=3.5in}}
\vskip 0.5cm
\caption{Histogram of the failure rates in identifying the correct H/P
class of the 20 amino acids obtained with method 3.}
\label{fig:f}
\end{figure}

In equation (\ref{eqn:en3}), the amplitudes of the two potential
wells, $V_1$ and $V_2$, were chosen to be of the order of 100 (the
energy unit is $| \epsilon(1,1)|=1$). By using the strategy described in
subsection B we obtained the best results for $\epsilon(1,0) =0.285509
$ and $\epsilon(0,0) =1.520444$, for which the design success rate was
73.1\%, slightly lower than for the previous attempt. The detailed
results are summarized in Figs.  \ref{fig:e} and \ref{fig:f}. Once
more, the failure rates on individual amino acids differ appreciably
from previous attempts (the amino acid failure rates are, however,
robust against small changes in the optimized parameters).

An artificial way of increasing the design success would be to include
alanine in the class of polar residues. This would make the success
rate of method 1 grow to $73.7\%$ . For the second method the success
increase is, {\em a priori\/} not as easy to estimate. In fact, it can
be expected that changing the class of alanine could trigger a cascade
effect modifying the overall success score.

Surprisingly this was not the case; by changing alanine from $H$ to
$P$ the best overall success grew to $75.1\%$. This is almost exactly
the value that one would have obtained by taking, for alanine, the
complementary of the failure rate in Fig. \ref{fig:d}, suggesting an
apparently weak correlation between the position of alanine and other
residues.

\subsection{Structural homology}

We complete the discussion of our design attempts by considering the
case of protein homology. It is well-known \onlinecite{b0,b1,b2} that there
is no strict 
one-to-one correspondence between structures and sequences; indeed, it
is often the case that native states of proteins whose mutual identity
is around 80\% fold into nearly the same structure (the RMSD being
typically less than 0.8 \AA\ per residue).  In this case the two
proteins are said to be homologous. A thorough check of a design
procedure should then allow for the possibility that the designed
sequence, $S^\prime$, is homologous to the target one, $S$. This can
conveniently be done by comparing $S^\prime$ with all the known
sequences homologous to $S$.

Table \ref{tab:omo} lists, for each of our 20 target structures, the
names of proteins with the same length and high structural identity
(this includes, of course, the true target sequence). It can be
noticed that, in a number of cases, it was not possible to identify
any homologous sequence.

\noindent We carried out the same design procedure as in method 3, but
with the following proviso: in case a target structure is associated
with more than one sequence in Table \ref{tab:omo}, we measure the
design success against each sequence in this set and then take the
highest value as the design score.

It turned out that, for nearly all cases, the highest sequence
identity was attained with the true target sequence, not the
homologous ones. For this reason the overall success was 73.4\%, very
close to the previous result.

The inability to improve significantly the success score over the four
attempts discussed above is, possibly, suggestive that important
features of real proteins have been mistreated.

One possibility (K. A. Dill, private communication) is that naturally
occurring proteins may not have necessarily evolved to maximize the
occupation probability (\ref{eqn:occprob}). According to this point of
view, designed sequences may differ from natural ones because they
have a higher thermodynamic stability.

Another possible explanation for our inability to correctly recognize
as many as 25\% of the residues may be the coarse graining of the 20
types of amino acids into just two classes, H and P.  In the following
section we will try to explore this possibility by studying a lattice
model for proteins. Indeed it will appear that the HP coarse graining
seriously limits the maximum achievable success rate.

\section{Disadvantages of the HP coarse-graining}

In this section we want to highlight the fact that an H-P
coarse-graining can partially mislead the design procedure. To
substantiate this claim we perform a lattice test in which proteins
are schematically represented by self-avoiding walks on a square
lattice. The amino acids are located at the sites touched by the
chains and they are divided in four types, according to their
different degree of hydrophobicity or polarity: $H_1$, $H_2$, $P_1$,
$P_2$. Two non-consecutive amino acids will be said to be in contact
if they are spaced only one lattice unit apart.  The contact
potentials, $u(\sigma_i, \sigma_j)$ (where $\sigma \in
\{H_1,H_2,P_1,P_2\}$) between two interacting sites were chosen to be

\begin{equation}
u = \left(
\begin{array}{cccc}
-2 & - {3 \over 2} & -{2 \over 3} & -{1 \over 2} \\
- {3 \over 2} & -1 & -{1 \over 2} & -{1 \over 3} \\
- {2 \over 3} & -{1 \over 2} & -{1 \over 2} &  -{3 \over 8} \\
- {1 \over 2} & -{1 \over 3} & -{3 \over 8} &  -{1 \over 4}
\end{array}
\right) \ .
\label{eqn:matu}
\end{equation}

\noindent The entries of the matrix $u$ were chosen so that the
residues $H_1$ and $H_2$ are energetically favoured to stay in the
interior of the protein, unlike $P_1$ and $P_2$.  We consider chains
made of 12 residues, a case where an exhaustive enumeration of all
possible conformations can be carried out with modest numerical
effort. Thus, for a given sequence of amino acids $S= \{\sigma_1,
\sigma_2, ... , \sigma_{12}\}$, we can verify if it has a unique
ground state (i.e. if it is a stable protein) or not.  We then
generated and stored a set of 1000 sequences $S_1, S_2 , ... ,
S_{1000}$ with unique ground state conformations, $\Gamma(S_i)$. Our
aim was to verify whether a coarse graining of the original 4 types of
amino acids into just two HP classes still allows a correct design
over the 1000 conformations. As a first attempt, we gathered the amino
acids $H_1$ and $H_2$ in class H and $P_1, P_2$ in class P and
introduced a new set of interaction potentials, $\tilde{u}(\sigma_i,
\sigma_j)$ between these two new classes by averaging out the
corresponding 2x2 H-P blocks in the matrix (\ref{eqn:matu}),

\begin{equation}
\tilde{u}=
\left(
\begin{array}{cc}
- {3 \over 2} & -{1 \over 2} \\
- {1 \over 2} & -{3 \over 8} \\
\end{array}
\right)\ .
\end{equation}

\begin{figure}[htbp]
\centerline{\psfig{figure=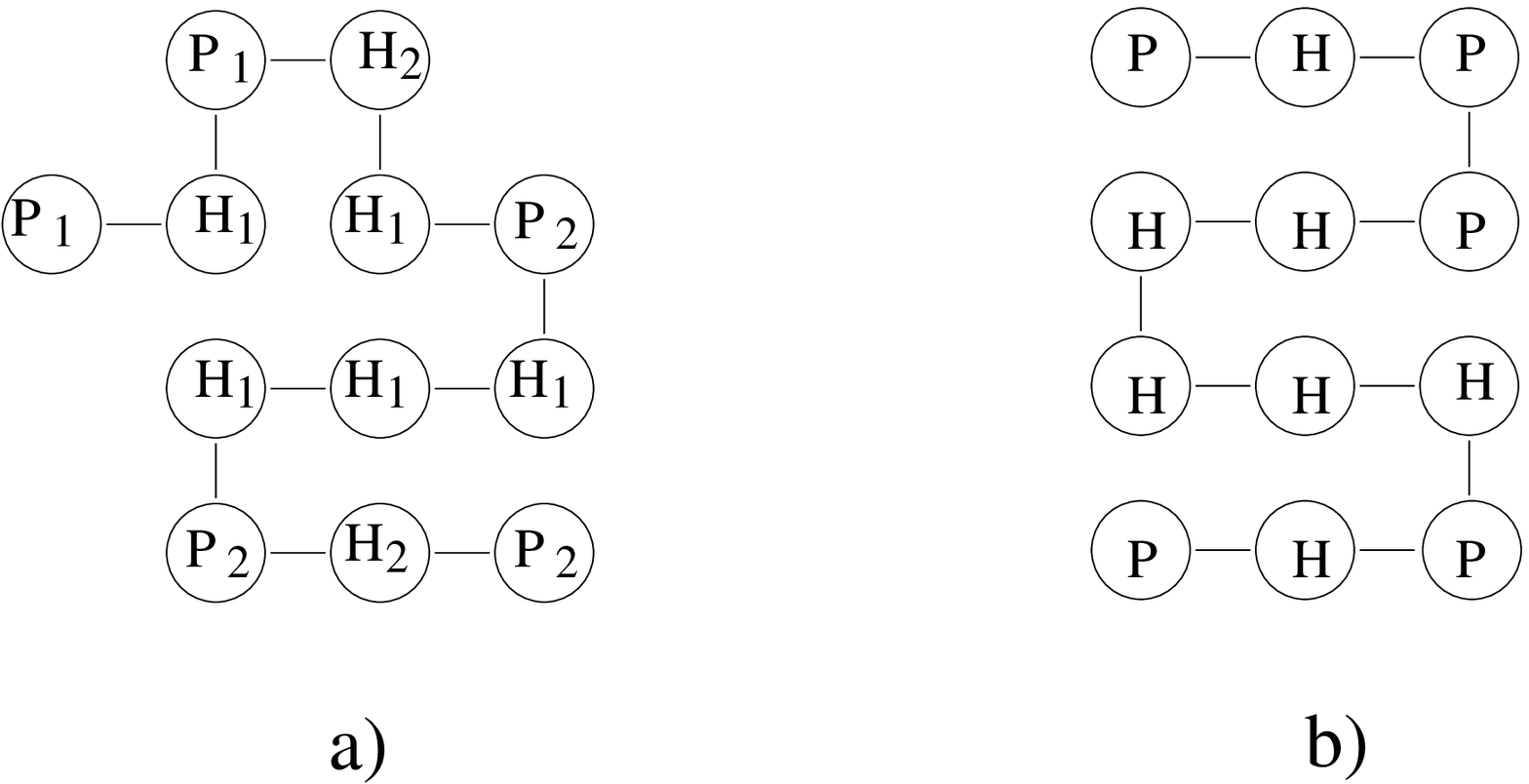,width=3.5in}}
\caption{a) The unique ground state conformation for sequence $S=\{
P_1, H_1, P_1, H_2, H_1, P_2, H_1, H_1, H_1, P_2, H_2, P_2 \}$. The
ground state energy of $S$ is -8. The HP coarse graining of $S$
yields a sequence, $\tilde{S}=\{P, H, P, H, H, P, H, H, H, P, H, P
\}$ which has energy -6.5 on conformation (a). The true ground state
of $\tilde{S}$ has energy -6.875 and is represented in (b).}
\label{fig:walk}
\end{figure}

\noindent Then we transformed each sequence $S_i$ in the corresponding
coarse grained form, $\tilde{S}_i$ and by means of exact enumeration
we tried to verify if $\tilde{S}_i$ still had a unique state of lowest
energy (adopting the coarse grained interactions) on conformation
$\Gamma(S_i)$.  Disappointingly, only in 718 cases was the answer
affirmative. A case where the procedure fails is shown in
Fig.~\ref{fig:walk}.

To make our test more stringent, we repeated the procedure by setting
\begin{eqnarray}
\tilde{u}(H,H) &=& -1,\\
\tilde{u}(H,P) &=& \tilde{u}(H,P)=z \\
\tilde{u}(H,H) &=&y
\end{eqnarray}

\noindent and optimally varied $y$ and $z$ in order to maximize the
number of ground states correctly predicted. We have found an optimal
solution for $z=y=0.86$ for which 830 of the original 1000
conformations were identified as native states of the coarse-grained
sequences.

Although one cannot extend the same quantitative study of the effects
of the HP coarse graining on real protein design, the results given
above suggest that not only they can be non negligible, but they can
significantly reduce the maximum design success that can be achieved.

\section{Summary}

To summarize we have presented several protein design methods for
identifying the correct hydrophobic or polar (H/P) class of residues
on a set of 20 proteins. The simplest of these techniques, which is
readily implemented and takes nearly no CPU time, gave a success rate
of 73\%.  More sophisticated methods, encompassing negative design
features, failed to improve appreciably upon the previous attempt.

\noindent We have presented evidence showing that the highest failure
rate was found on residues with high secondary-structure
functionality. On the basis of numerical results on exactly solvable
models, it was also suggested that the HP coarse-graining may
significantly impair the design algorithms, thus possibly accounting
for the failure to increase the design success score by using
algorithms of growing complexity.

\acknowledgments{This work was supported in part by INFN sez. di
Trieste, NASA, NSF, NATO and the Center for Academic Computing at Penn
State. We are indebted to Anna Tramontano and Ruxandra Dima for useful
discussions.}

\newpage
\begin{table}[htbp]
\begin{center}
\begin{tabular}[h]{| l | l| l| } \hline
Label  &  PDB File  & Len\\ \hline \hline
1  & 1bba  &   36 \\
2  & 1bbl  &   37 \\
3  & 3ebx  &   62 \\
4  & 1aba  &   87\\
5  & 2hpr  &   87\\
6  & 1aps  &   98\\
7  & 1aaj  &  105\\
8  & 1erv  &  105\\
9  & 1ycc  &  108\\
10 & 5cpv  &  108\\
11 & 3rn3  &  124\\
12 & 1hel  &  129\\
13 & 1ifb  &  131\\
14 & 1ecd  &  136\\
15 & 1osa  &  148\\
16 & 1mbd  &  153\\
17 & 1ra8  &  159\\
18 & 1l92  &  162\\
19 & 2lzm  &  164\\
20 & 9pap  &  212\\
\end{tabular}
\end{center}
\vskip 0.5cm
\caption{The names of the 20 proteins taken from the Protein Data Bank
and used in out design studies are given in column two. The number
of amino acids in each protein is given in the last column.}
\label{tab:files}
\end{table}

\begin{table}[htbp]
\begin{center}
\begin{tabular}{| l | l | l |}\hline \\
Amino Acid & Type & Freq. (\%)\\
\hline \hline
ALA & H &  8.85\\
VAL & H &  6.59\\
LEU & H &  6.85\\
ILE & H &  5.36\\
CYS & H &  2.04\\
MET & H &  2.42\\
PHE & H &  4.38\\
TYR & H &  3.32\\
TRP & H &  1.40\\
GLY & P &  7.91\\
PRO & P &  3.23\\
HIS & P &  2.17\\
SER & P &  5.70\\
THR & P &  5.87\\
LYS & P &  8.08\\
ARG & P &  4.64\\
ASP & P &  5.96\\
ASN & P &  5.15\\
GLU & P &  6.76\\
GLN & P &  3.32
\end{tabular}
\end{center}
\vskip 0.5cm
\caption{Following Sun {\em et al.} {\protect{\onlinecite{i16}}} we
distributed the 20 amino acids in H/P classes as in the second
column. The third column gives the relative frequency of a given
amino acid in the 20 proteins of Table \protect{\ref{tab:files}}.}
\label{tab:hp}
\end{table}

\newpage
\begin{table}[htbp]
\begin{center}
\begin{tabular}{| l | l | }\hline \\
Target structure & Homologous Sequences \\
\hline \hline
1bba  &  1bba \\
1bbl  &  1bbl\\
3ebx  &  3ebx, 5ebx, 1nxb \\
1aba  &  1aba\\
2hpr  &  2hpr\\
1aps  &  1aps\\
1aaj  &  1aaj\\
1erv  &  1erv, 1eru, 3trx, 1trw\\
1ycc  &  1ycc, 1csu, 1raq, 1crj, 1cig\\
5cpv  &  5cpv\\
3rn3  &  3rn3\\
1hel  &  1hel\\
1ifb  &  1ifb\\
1ecd  &  1ecd\\
1osa  &  1osa, 3cln, 4cln, 1cll\\
1mbd  &  1mbd\\
1ra8  &  1ra8, 1jom\\
1l92  &  1l92, 7lzm, 1l74, 1l72, 1l64\\
2lzm  &  2lzm\\
9pap  &  9pap\\
\end{tabular}
\end{center}
\vskip 0.5cm
\caption{For each protein in the original set (first column) we
identified, when possible, a set of proteins with high structural
identity (second column). The outcome of our design procedure was then
checked against the set of structurally homologous sequences.}
\label{tab:omo}
\end{table}

\newpage

\begin{references}

\bibitem{i14}  Bowie, J. U, Luthy, R. \& Eisenberg, D., (1991),
A method to identify protein sequences that fold into a known 3-
dimensional structure.
{\em Science} {\bf 253}, 164-170

\bibitem{b0} Branden, C. \& Tooze, J. (1991) in {\em Introduction to protein
structure}, Garland Publishing, New York

\bibitem{i1} Cordes, M. H. J., Davidson, A. R. \& Sauer, R. T. (1996),
Sequence space, folding and protein design.
{\em Curr. Opin. in Struct. Biol.} {\bf 6}, 3-10

\bibitem{b1} Creighton, T.E. (1983) in {\em Proteins: structures and
molecular properties}, W. H. Freeman ed., New York

\bibitem{i10} Deutsch, J. M. \& Kurosky, T. (1996),
New algorithm for protein design.
{\em Phys. Rev. Lett.} {\bf 76}, 323-326.

\bibitem{b2} Fletterick, R. J., Schroer, T., \& Matela, R. J. (1985) in {\em
Molecular structure: macromolecules in three dimensions},
Blackwell Scientific, Oxford, UK

\bibitem{i5} Kamtekar, S., Schiffer, J. M., Xiong, H.,
Babik, J. M. \& Hecht, M. H. (1993),
Protein design by binary patterning of polar and nonpolar amino-acids.
{\em Science} {\bf 262}, 1680-1685

\bibitem{i2} Kuroda, Y., Nakai, T. \& Ohkubo, T. (1994), Solution
structure of a de-novo helical protein by 2d-NMRr spectroscopy. {\em
J. Mol. Biol.} {\bf 236}, 862-868

\bibitem{i15a}  Lau, K. F. \&  Dill, K. A. (1989)
A lattice statistical-mechanics model of the conforma,tional and
sequence-spaces of proteins. {\em Macromolecules} {\bf 22}, 3986-3997

\bibitem{i4} Lombardi, A., Bryson, J. W. \&  DeGrado W. F. (1997),
De novo design of heterotrimeric coiled coils
{\em Biopolym. Peptide Sci.} {\bf40}, 495-504

\bibitem{i17} Maiorov, V. N. \& Crippen, M. G. (1992), Contact potential that
recognizes the correct folding of globular proteins.
{\em J. Mol. Biol.}{\bf 227}, 876-888

\bibitem{i18} Micheletti, C., Seno, F., Maritan, A. \& J. R. Banavar (1997),
{\em Protein Design in a Lattice Model of Hydrophobic and
Polar Amino Acids} preprint

\bibitem{i12} Morrissey, M. P. \& Shakhnovich, E. I. (1996),
Design of proteins with selected thermal properties
{\em Folding and Design} {\bf 1}, 391-405

\bibitem{i15}  Pabo, C. (1983), Designing proteins and peptides.
{\em Nature} {\bf 301}, 200

\bibitem{i13}  Ponder, J. W. \& Richards, F. M. (1987),
Tertiary templates for proteins - use of packing criteria in the
enumeration of allowed sequences for different structural classes.
{\em J. Mol. Biol.} {\bf 193}, 775-791

\bibitem{i3} Quinn, T. P., Tweedy, N. B., Williams R. W., Richardson,
J. S. \& Richardson, D. C. (1994),
Beta-doublet - de-novo design, synthesis, and characterization of a
beta-sandwich protein.
{\em Proc. Natl. Acad. Sci USA} {\bf 91}, 8747-8751

\bibitem{i11}  Seno, F.,  Vendruscolo, M.,  Maritan, A., \& Banavar,
J. R. (1996), Optimal protein design procedure.
{\em Phys. Rev. Lett.} {\bf 77}, 1901-1904

\bibitem{i8}  Shakhnovich, E. I. (1994),
Proteins with selected sequences fold into unique native conformation.
{\em Phys. Rev. Lett.}  {\bf 72}, 3907-3910

\bibitem{new} Shakhnovich, E. I. \& Gutin A. M., (1993),
A new approach to the design of stable proteins.
{\em Protein Engineering} {\bf 6}, 793-800 

\bibitem{i16}  Sun, S. J., Brem, R., Chan, H. S. \& Dill, K. A. (1995),
Designing amino-acid-sequences to fold with good hydrophobic cores.
{\em Protein Engineering} {\bf 8}, 1205-1213

\bibitem{i9}  Yue, K., \& Dill, K. A. (1992)
Inverse protein folding problem - designing polymer sequences.
{\em Proc. Natl. Acad. Sci. USA} {\bf 89}, 4163-4167

\bibitem{i6} Yue, K.,  Fiebig, K. M.,  Thomas, P. D.,  Chan, H. S.,
Shackhnovich, E. I.,  \& Dill, K. A., (1995),
A test of lattice protein-folding algorithms.
{\em Proc. Natl. Acad. Sci. USA} {\bf 92}, 325-329 (1995)

\end{references}
\end{document}